\documentstyle[12pt]{article}
\newcommand{\sptwo}{1.4}

\newcommand{\doublespace}{\edef\baselinestretch{\sptwo}\Large\normalsize}

\begin{document}
\doublespace

\begin{center}
{\bf Beyond the Thomas-Fermi Approximation for Nonlinear Dynamics
of Trapped Bose-Condensed Gases}\\
\renewcommand\thefootnote{\fnsymbol{footnote}}
{Alexander L. Zubarev\footnote{ e-mail: zubareva$@$physics.purdue.edu} and
Yeong E. Kim \footnote{ e-mail:yekim$@$physics.purdue.edu}}\\ 
Department of Physics, Purdue University\\
West Lafayette, Indiana  47907\\
{\bf Abstract}
\end{center}
\begin{quote}
We present an analytical approximation for nonlinear dynamics of trapped Bose-condensed gases. The new approximation is a substantial improvement over the Thomas-Fermi 
 approximation and is shown to be applicable for systems with a rather small 
number of atoms $N$. The calculated aspect ratios after ballistic expansion are found to be in good agreement with those observed in recent experiments.
\end{quote}

\vspace{5mm}

\vspace{55 mm}
\noindent

\pagebreak
The newly created Bose-Einstein condensates (BEC) of weakly interacting
alkali-metal atoms [1] 
 stimulated a large number of theoretical investigations (see recent reviews [2]). A mean-field approach, based on the time-dependent Gross-Pitaevskii (GP)
equation [3], is the most widely used theory for nonlinear dynamics of trapped Bose-condensed gases at zero temperature.

In the limit of a large number  of atoms $N$,  
the determination of the condensate wave function is 
simplified by neglecting the kinetic energy term.
This approximation is known as the Thomas-Fermi (TF) approximation [4,5].
It has been  used quite 
extensively [7-13], including the explicit time evolution of the condensate (
shape of profiles, aspect ratio etc.) during the expansion after switching off 
the trap [7,9].
We note that the validity of the TF approximation depends not only on the number of
atoms $N$, but also depends on properties of the traps.

In this letter we develop an analytical approximation which 
is a substantial improvement over 
 the TF results for rather small number of atoms $N$. 
The aspect ratios calculated from the new approximation method are also found to be in good agreement with those 
observed in recent experiments [14].

In the mean-field approximation, the ground state energy of the system is given by the Ginzburg-Pitaevskii-Gross (GPG) energy functional [3,15]
$$
\frac{E}{N}=<\Psi\mid\sum_{i=1}^3 H_i\mid \Psi>+\frac{g N}{2}\int\mid\Psi\mid^4 d\vec{r},
\eqno{(1)}
$$
with 
$$
H_i=-\frac{\hbar^2}{2m}\frac{\partial^2}{\partial x_i^2}+\frac{m\omega_i^2}{2} x_i^2,
\eqno{(2)}
$$
and $g=4\pi\hbar^2 a/m$, where $a$ is the $S$-wave scattering length, $N$ is the 
number of atoms in the BEC, and $\Psi$ is the condensate wave function normalized as $\int \mid\Psi\mid^2 d\vec{r}=1$.

 We introduce an auxiliary Hamiltonians
$$
\tilde{H}_i=\frac{\hbar \omega_i}{2}\sqrt{\gamma_i}+\frac{m \omega_i^2}{2}(1-\gamma_i)x_i^2,
\eqno{(3)}
$$
where $\gamma_i$ are parameters, $0\leq\gamma_i<1$,
and rewrite Eq.(1) as
$$
\frac{E}{N}=<\Psi\mid\sum_{i=1}^3(H_i-\tilde{H}_i)\mid\Psi>+
<\Psi\mid\sum_{i=1}^3\tilde{H}_i\mid \Psi>+\frac{g N}{2}\int\mid\Psi\mid^4
d\vec{r}.
\eqno{(4)}
$$
 Omission of $<\Psi\mid\sum_{i=1}^3(H_i-\tilde{H}_i)\mid\Psi>$ in Eq.(4)
yields our approximation for the ground state,
$$
\frac{E}{N}=\sum_{i=1}^3\frac{\hbar \omega_i}{2}\sqrt{\gamma_i}+\frac{E_{TF}
(\tilde{\omega}_x,\tilde{\omega}_y,\tilde{\omega}_z)}{N},
\eqno{(5)},
$$
$$
\mu=\sum_{i=1}^3\frac{\hbar \omega_i}{2}\sqrt{\gamma_i}+\mu_{TF}(\tilde{\omega}_x,\tilde{\omega}_y,\tilde{\omega}_z),
\eqno{(6})
$$
and
$$
\rho(\vec{r})=\rho^{TF}(\tilde{\omega}_x,\tilde{\omega}_y,\tilde{\omega}_z,\vec{r}),
\eqno{(7)}
$$
where $\tilde{\omega}_i=\sqrt{1-\gamma_i}\omega_i$.  $E_{TF}, \mu_{TF}$, and $\rho^{TF}$ are the Thomas-Fermi energy, chemical potential, and density, respectively, which are given by
$$
E_{TF}(\omega_x, \omega_y, \omega_z)=\frac{5}{14}N[\frac{15}{4 \pi}\omega_x \omega_y \omega_z
m^{3/2} g N]^{2/5},
\eqno{(8)}
$$
 $$
\mu_{TF}(\omega_x, \omega_y, \omega_z)=\frac{1}{2} [\frac{15}{4 \pi}\omega_x 
\omega_y \omega_z
m^{3/2} g N]^{2/5},
\eqno{(9)}
$$
and
$$
\rho^{TF}(\omega_x, \omega_y, \omega_z,\vec{r})=
\frac{\mu_{TF}(\omega_x, \omega_y, \omega_z)}{Ng} (1-\sum_{i=1}^3
(\frac{x_i}{R_i^{TF}})^2) \theta(1-\sum_{i=1}^3(\frac{x_i}{R_i^{TF}})^2),
\eqno{(10)}
$$
where 
$$
(R_i^{TF})^2=\frac{2 \mu_{TF}(\omega_x, \omega_y, \omega_z)}{m\omega_i^2}.
\eqno{(11)}
$$

Projecting $\mid\Psi>$ on the complete basis states  $\mid n>$, obtained from $h_i=-\frac{\hbar^2}{2 m}\frac{\partial^2}{x_i^2}+\frac{m \omega_i^2 \gamma_i}{2}x_i^2$, and $h_i \mid n>=\epsilon_n \mid n>$, we get
$$
<\Psi\mid h_i \mid \Psi>=\sum_{n} \epsilon_n \mid<\Psi\mid n><n\mid \Psi>\mid \:
\geq\epsilon_1=\frac{\hbar \omega_i\sqrt{\gamma_i}}{2}.
\eqno{(12)}
$$
Therefore,
we conclude that our approximation for energy, given by Eq.(5), is a lower bound to the ground 
state energy, Eq.(4). Therefore a set of the optimal values of parameters $\gamma_i$ which 
maximizes the energy, Eq.(5),
will yield an optimal value for the ground-state energy given by
$$
\frac{E}{N}=\max_{\gamma_x,\gamma_y,\gamma_z}[\sum_{i=1}^3\frac{\hbar \omega_i}{2}\sqrt{\gamma_i}+\frac{5}{14}(\frac{15}{4\pi} gN m^{3/2}\prod_{i=1}^3((1-\gamma_i)\omega_i))^{2/5}].
\eqno{(13)}
$$
Since the TF approximation corresponds to the case of $\gamma_i=0$, we have
$$ 
E_{TF}\leq E\leq E_{exact},
\eqno{(14)}
$$
where $E_{exact}$ is the exact mean-field energy.

To study the validity of our approximation, we consider 
 an example of the  ground state of $^{87}Rb$ atoms in a harmonic trap, 
as investigated  in Ref.[16] with the S-wave
triplet-spin scattering length $a=100 a_B$, where $a_B$ is the Bohr radius, the axial frequency $\omega_z/2 \pi= 220$Hz, and assymetry parameter 
$\lambda=\omega_z/\omega_{\perp}=\sqrt{8}$, 
where $\omega_x=\omega_y=\omega_{\perp}$.

Using our approximation, we calculate
 the  energy per particle, $E/N$, the
chemical potential  $\mu$, and the average transverse sizes $\sqrt{\overline{x^2}}$ 
and 
vertical  sizes$\sqrt{\overline{z^2}}$, using Eqs. (5-6,13). 
The calculated results are compared 
 with those obtained from  the numerical solutions of the GP equation, 
$E_{num}/N$, $\mu_{num}$,
$\sqrt{\overline{x^2}_{num}}$ and $\sqrt{\overline{z^2}_{num}}$ [16]  in Table I, and with 
  those obtained  in the TF approximation $E_{TF}/N$, $\mu_{TF}$,
$\sqrt{\overline{x^2}_{TF}}$ and $\sqrt{\overline{z^2}_{TF}}$ in Table II.
These comparisons 
show that our analytical approximation greatly improves the TF results for a
rather small  number $N$. For $100\leq N \leq 20000$, the difference between 
our results and those of the numerical solution of the GP equation [16] are less than 3\%.

Let us now  turn to application of our approximation for the time-dependent problems.
Consider the BEC  with the time-dependent harmonic potential 
$ V_t=(m/2) \sum_{i=1}^3\omega_i^2(t) x_i^2$.

In Ref.[7] the following anzatz
$$
\Psi(\vec{r},t)=
 \frac{\Phi(x_1/\lambda_1,x_2/\lambda_2,x_3/
\lambda_3,t)}{\sqrt{\lambda_1 \lambda_2 \lambda_3}}exp[-i\beta(t)+im\sum_{i=1}^3
\frac{x_i^2}{2 \hbar}
 \frac{\dot{\lambda}_i}{\lambda_i}]
\eqno{(15)}
$$
has been used for the solution of the time-dependent GP equation
$$
i\hbar\frac{\partial\Psi}{\partial t}=-\frac{\hbar^2}{2m}\Delta \Psi+V_t(\vec{r}
,t) \Psi+
Ng\mid\Psi\mid^2\Psi
\eqno{(16)}
$$
 (a similar treatment for isotropic traps has been developed in Ref.[17]). 
We choose
 $\beta$ and $\lambda$ satisfying the following equations
$$
\hbar \dot{\beta}=\sum_{k=1}^3\frac{\hbar \omega_k\sqrt{\gamma_k}}{2 \lambda_k^2
}-\frac{\mu_{TF}(\tilde{\omega}_x, \tilde{\omega}_y,\tilde{\omega}_z)}{\lambda_1
 \lambda_2 \lambda_3},\: \beta(0)=0,
\eqno{(17)}
$$
and
$$
\ddot{\lambda}_k\lambda_k+\lambda_k^2 \omega_k^2(t)-\frac{\omega_k^2\gamma_k}
{\lambda_k^2}=\frac{(1-\gamma_k)\omega_k^2}{\lambda_1 \lambda_2 \lambda_3},\:
\lambda_k(0)=1,\: \dot{\lambda}_k(0)=0,
\eqno{(18)}
$$
with
 $\omega_k=\omega_k(0)$, and $\tilde{\omega}_k=\sqrt{1-\gamma_k} \omega_k$.
For a special case of $\gamma_k=0$, the above choices of $\beta$ and $\lambda_k$ reduce to those of Ref.[7]. With substitutions of Eqs.(15,17-18), Eq.(16) becomes 
$$
i\hbar\frac{\partial \Phi}{\partial t}=\sum_{k=1}^3 \frac{H_k-\tilde{H}_k}{
\lambda_k^2
}\Phi+\frac{1}{\lambda_1 \lambda_2 \lambda_3}[-\mu_{TF}(\tilde{\omega_x},
\tilde{\omega_y},\tilde{\omega}_z)+\frac{m}{2}\sum_{k=1}^3 x_k^2 \tilde{\omega}_
k^2+
N g \mid \Phi \mid^2]\Phi,
\eqno{(19)}
$$
with the initial condition $\Phi(\vec{r},0)=\Psi(\vec{r},0)$,
where $\Psi(\vec{r},0)$ is a solution of the time-independent mean-field equation
$$
-\frac{\hbar^2}{2m}\Delta \Psi(\vec{r},0)+V_t(\vec{r},0)\Psi(\vec{r},0)+
N g \mid \Psi(\vec{r},0)\mid^2 \Psi(\vec{r},0)=\mu \Psi(\vec{r},0).
\eqno{(20)}
$$

By neglecting $\sum_{k=1}^3\frac{H_k-\tilde{H}_k}{\lambda_k^2}$ in Eq. (19),
we obtain a generalization of our approximation, Eqs.(5-7,13) to the time-dependent problem
$$
\mid\Psi(\vec{r},t)\mid^2=\frac{\rho_{TF}(\tilde{\omega}_x,\tilde{\omega}_y,\tilde{\omega}_z,x_1/\lambda_1(t),x_2/\lambda_2(t),x_3/\lambda_3(t))}{\lambda_1(t)
\lambda_2(t) \lambda_3(t)},
\eqno{(21)}
$$
where all the dynamics is in the evolution of the scaling parameters $\lambda_k$,
Eq.(18).

For the case $\omega_x=\omega_y=\omega_{\perp}$ and $\lambda_1=\lambda_2=\lambda_{\perp}$, the aspect ratio of the cloud  in our approximation is given by
$$
R(t)=\sqrt{\frac{\overline{x^2}(t)}{\overline{z^2}(t)}}=
\frac{\omega_z \lambda_{\perp}(t) \sqrt{1-\gamma_z}}{\omega_{\perp} \lambda_z(t)\sqrt{1-\gamma_{\perp}}}.
\eqno{(22)}
$$

As an example,
 we consider application of the above results to the experimental data with $^{23}Na$ atoms obtained in the Ioffe-Pritchard type magnetic trap with radial and axial trapping frequences of $\omega_{\perp}/(2 \pi)=360$ Hz and $\omega_z/(2 \pi)=3.5$ Hz [14], respectively. In our analysis  we use $a=2.75$nm, $t=4$ ms, and $a/a_{\perp}=2.488\times 10^{-3}$, where $a_{\perp}=\sqrt{\hbar/m\omega_{\perp}}$. As in Ref. [7] we consider a sudden and total opening of the trap at $t=0$. For this case Eq. (17) becomes
$$
\frac{d^2 \lambda_{\perp}}{d \tau^2}=\frac{\gamma_{\perp}}{\lambda_{\perp}^3}+\frac{1-\gamma_{\perp}}{\lambda_{\perp}^3\lambda_z},
\eqno{(23)}
$$
$$
\frac{d^2 \lambda_z}{d\tau^2}=(\frac{\gamma_z}{\lambda_z^2}+\frac{1-\gamma_z}{\lambda_{\perp} \lambda_z})\epsilon^2,
\eqno{(24)}
$$
where $\tau=\omega_{\perp} t$ and $\epsilon=\omega_z/\omega_{\perp}\ll1$.

To the zero-th order in $\epsilon$, we have  $\lambda_z=1$ and $\lambda_{\perp}=\sqrt{1+\tau^2}$. For the experimental conditions [14], the terms in $\epsilon^2$ are negligible.

In Table III, we give the calculated values of the aspect ratio $R(t)$ of the $^{23}Na$ atoms cloud, after ballistic expansion of $t=4$ ms, and the calculated values of parameters $\gamma_{\perp} $ and $\gamma_z$ which we fix from Eq.(13) ,
with $\omega_i=\omega_i(0)$.
 One can easily see that the TF approximation is valid ($\gamma_z\approx 0$) along the
long axis of the cloud, but not in the radial direction ($\gamma_{\perp}\neq 0$), as
 has been noted already in Ref. [18].

Our calculated results for $R(t)$ are compared with the recent experimental data [14] in Figure 1.
 This comparison shows that our predictions for the aspect ratio $R(t)$ are in good agreement with experimental data obtained by the MIT group [14].

In conclusion, we have developed an analytical approximation which provides a substantial improvement over the TF approximation  for nonlinear dynamics of trapped Bose-condensed gases for a rather small number of atoms $N$. The approximation is very useful since it provides an easy quantitive tool for the analysis of experiments on trapped condensed gases.

We are grateful to the group at MIT for providing us with the experimental data.  
\pagebreak

Table I. Comparison of the results of our approximation  for the ground state of $^{87}Rb$ atoms, calculated from
 Eqs.(5-6,13) and those obtained from the numerical solutions of the GP equation [16]. Chemical potential and energy are in units of $\hbar \omega_{\perp}$, and
length is in units of $\sqrt{\hbar/m \omega_{\perp}}$

\vspace{8pt}
\begin{tabular}{lllllllll}
\hline\hline
$N$
&$E/N$
&$\mu$
&$\sqrt{\overline{x^2}}$
&$\sqrt{\overline{z^2}}$
&$E_{num}/N$
&$\mu_{num}$
&$\sqrt{\overline{x_{num}^2}}$
&$\sqrt{\overline{z_{num}^2}}$\\ \hline
100
&2.63
&2.82
&0.78
&0.43
&2.66
&2.88
&0.79
&0.44\\ \hline
200
&2.80
&3.13
&0.83
&0.45
&2.86
&3.21
&0.85
&0.45\\ \hline
500
&3.22
&3.82
&0.94
&0.47
&3.30
&3.94
&0.96
&0.47\\ \hline
2000
&4.49
&5.49
&1.21
&0.53
&4.61
&5.93
&1.23
&0.53\\ \hline
5000
&5.99
&8.00
&1.46
&0.59
&6.12
&8.14
&1.47
&0.59\\ \hline
10000
&7.63
&10.4
&1.68
&0.65
&7.76
&10.5
&1.69
&0.65\\ \hline
15000
&8.84
&12.1
&1.83
&0.69
&8.98
&12.2
&1.84
&0.70\\ \hline
20000
&9.84
&13.5
&1.94
&0.72
&9.98
&13.7
&1.94
&0.73
 \\ \hline\hline
\end{tabular}\\

\vspace{18pt}

Table II. Results of the TF approximation for the same case as Table I.

\vspace{8pt}

\begin{tabular}{lllll}
\hline\hline
$N$
&$E_{TF}/N$
&$\mu_{TF}$
&$\sqrt{\overline{x_{TF}^2}}$
&$\sqrt{\overline{z_{TF}^2}}$\\ \hline
100
&1.44
&1.60
&0.68
&0.24\\ \hline
200
&1.51
&2.11
&0.78
&0.27\\ \hline
500
&2.18
&3.05
&0.93
&0.33\\ \hline
2000
&3.79
&5.31
&1.23
&0.44\\ \hline
5000
&5.47
&7.66
&1.48
&0.52\\ \hline
10000
&7.22
&10.1
&1.70
&0.60\\ \hline
15000
&8.49
&11.9
&1.84
&0.65\\ \hline
20000
&9.53
&13.3
&1.95
&0.69 \\ \hline\hline
\end{tabular}\\

\pagebreak

Table III. Calculated aspect ratio of the $^{23}Na$ atoms cloud, using Eq.(22), after a ballistic expansion of $t=4$ ms, as a function of $N$. $N$ is in units of $10^5$.

\vspace{8pt}
\begin{tabular}{llll}
\hline\hline
$N$
&$R(t)$
&$\gamma_{\perp}$
&$\gamma_z$\\ \hline
1.2
&0.110
&0.354
&0.0 \\ \hline
0.8
&0.117
&0.426
&0.0\\ \hline
0.4
&0.132
&0.551
&0.0\\ \hline
0.1
&0.183
&0.766
&0.0
 \\ \hline\hline
\end{tabular}\\

\vspace{18pt}

\pagebreak

\doublespace
\begin{picture}(471,342)(-20,0)
\thicklines
{\bf
\put(-28.6,-4.28){0}
\put(-40,60.72){0.05}
\put(-40,125.72){0.10}
\put(-40,190.72){0.15}
\put(-40,255.72){0.20}
\put(-10,0){\vector(1,0){410}}
\put(400,-4.7){$N/10^3$}
\put(0,305){\line(1,0){390}}
\put(390,0){\line(0,1){305}}
\put(0,-10){\vector(0,1){325}}
\put(-4.7,325){$R$}
\put(-10,65){\line(10,0){10}}
\put(-10,130){\line(10,0){10}}
\put(-10,195){\line(10,0){10}}
\put(-10,260){\line(10,0){10}}
\put(124,0){\line(0,-10){10}}
\put(248,0){\line(0,-10){10}}
\put(372,0){\line(0,-10){10}}
\put(-2.85714,-20){0}
\put(118.286,-20){40}
\put(242.286,-20){80}
\put(363.429,-20){120}}
\put(382.013,139.737){\circle*{10}}
\put(221.701,138.385){\circle*{10}}
\put(187.981,162.076){\circle*{10}}
\put(154.107,150.595){\circle*{10}}
\put(249.909,140.179){\circle*{10}}
\put(302.054,145.765){\circle*{10}}
\put(275.614,139.172){\circle*{10}}
\put(277.151,143.91){\circle*{10}}
\put(315.087,145.558){\circle*{10}}
\put(281.038,138.589){\circle*{10}}
\put(247.626,145.462){\circle*{10}}
\put(256.025,149.118){\circle*{10}}
\put(203.221,149.621){\circle*{10}}
\put(185.313,142,683){\circle*{10}}
\put(140.909,139.23){\circle*{10}}
\put(163.66,140.868){\circle*{10}}
\put(134.573,147.698){\circle*{10}}
\put(69.1818,162.815){\circle*{10}}
\put(97.6196,170.534){\circle*{10}}
\put(81.6608,198.784){\circle*{10}}
\put(43.5671,164.919){\circle*{10}}
\put(68.2759,188.86){\circle*{10}}
\put(97.8062,161.823){\circle*{10}}
\put(93.8885,164.896){\circle*{10}}
\put(40.9079,184.545){\circle*{10}}
\put(62.6684,184.616){\circle*{10}}
\put(69.8889,204.698){\circle*{10}}
\put(29.6619,214.14){\circle*{10}}
\put(26.9586,290.443){\circle*{10}}
\put(49.3932,187.94){\circle*{10}}
\put(82.6293,174.197){\circle*{10}}
\put(67.5602,191.705){\circle*{10}}
\put(58.1005,168.49){\circle*{10}}
\put(82.0595,159.058){\circle*{10}}
\thicklines
\put(0,114.4){\dashbox{10}(390,0){$~$}}
\put(26.959,246.509){$\diamond$}
\put(31,237.445){$\diamond$}
\put(46.5,213.898){$\diamond$}
\put(62,199.484){$\diamond$}
\put(77.5,189.497){$\diamond$}
\put(93,182.057){$\diamond$}
\put(108.5,176.245){$\diamond$}
\put(124,171.545){$\diamond$}
\put(155,164.35){$\diamond$}
\put(186,159.045){$\diamond$}
\put(217,154.937){$\diamond$}
\put(248,151.642){$\diamond$}
\put(279,148.929){$\diamond$}
\put(310,146.648){$\diamond$}
\put(341,144.699){$\diamond$}
\put(372,143.104){$\diamond$}
\put(382.013,141.225){$\diamond$}
\end{picture}

\vspace{10mm}

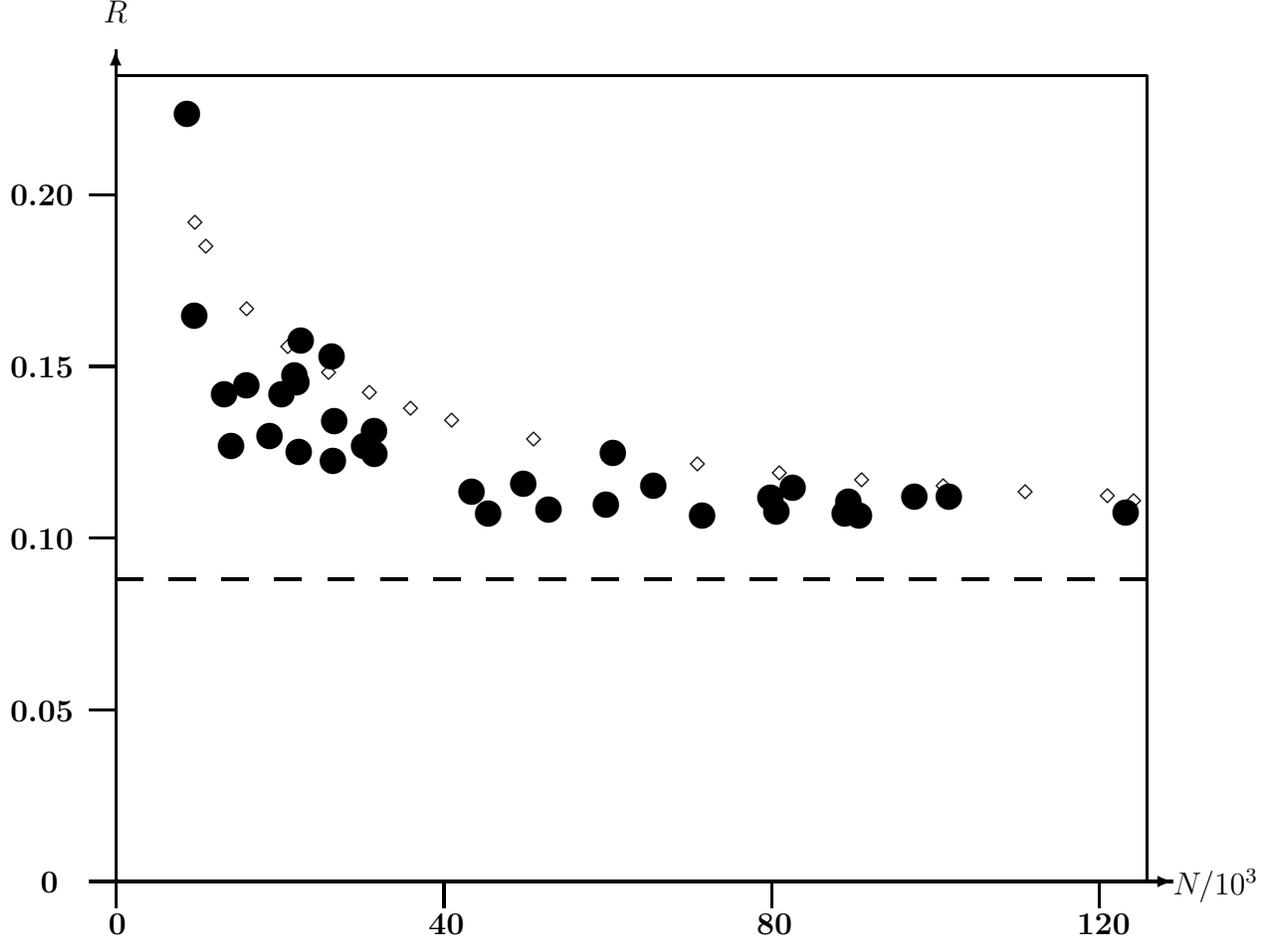
FIG. 1.  Aspect ratio $R$  of the $^{23}Na$ atom  cloud after a ballistic expansion of
$t=4$ms, as a function of the number of atoms $N$, with 
$\omega_{\perp}(0)=2 \pi \times
360$Hz, $\omega_z(0)=2 \pi \times$ 3.5 Hz.
Diamonds, dashed line, and circular dots represent the results of theoretical calculations using Eq.(22), the TF approximation, and  experimental data from
 MIT group [14], respectively.

\pagebreak

\begin{center}
{\bf References}
\end{center}

\vspace{8pt}
\noindent
1.  http://amo.phy.gasou.edu/bec.html; http://jilawww.colorado.edu/bec/
 and references therein.

\noindent
2. A.L. Fetter and A.A. Svidzinsky, J.Phys.: Condens. Matter { \bf13}, R135 (2001);
 K. Burnett, M. Edwards, and C.W. Clark, Phys. Today, {\bf 52}, 37 (1999);
 F. Dalfovo, S. Giorgini, L. Pitaevskii, and S. Stringari, Rev. Mod. Phys.
 {\bf 71}, 463 (1999); 
S. Parkins, and D.F. Walls, Phys. Rep. {\bf 303}, 1 (1998).

\noindent
3. L.P. Pitaevskii, Zh. Eksp. Teor. Fiz. {\bf 40}, 646 (1961) [Sov. Phys. JETP {\bf 13}, 451 (1961)];
   E.P. Gross, Nuovo Cimento {\bf 20}, 454 (1961); J. Math. Phys. {\bf 4}, 195 (1963).
 
\noindent
4. D.A. Huse and E.D. Siggia, J. Low. Temp. Phys. {\bf 46}, 137 (1982).

\noindent
5. G. Baum and  C.J. Pethick, Phys. Rev. Lett. {\bf 76}, 6 (1996).

\noindent
6. Tin-Lun-Ho and V.B. Shenoy, Phys. Rev. Lett. {\bf77}, 3276 (1996).

\noindent
7. Y. Castin and K. Dum, Phys. Rev. Lett. {\bf 77}, 5315 (1996).

\noindent
8. F. Dalfovo, L.P. Pitaevskii, and S. Stringari, Phys. Rev. A{\bf 54},
4213 (1996).

\noindent
9. F. Dalfovo, C. Minniti, S. Stringari, and L. Pitaevskii, Phys. Lett. A{\bf 227}, 259 (1977).

\noindent
10. E. Lundth, C.J. Pethik and H. Smith, Phys. Rev. A{\bf 55}, 2126 (1997).

\noindent
11.A.L. Fetter and D.L. Feder, Phys. Rev. A{\bf58}, 3185 (1998).

\noindent
12. E. Timmermans, P. Tommasini, and K. Huang, Phys. Rev. A{\bf 55}, 3645 (1997).

\noindent
13. Y.E. Kim and A.L. Zubarev, Phys. Rev. A{\bf64}, 013603 (2001).  

\noindent
14. A. G\"orlitz, J.M. Vogels, A.E.  Leanhardt, C. Raman, T.L. Gustavson,
J.R. Abo-Shaeer, A.P. Chikkatur, S. Gupta, S. Inouye, T.P. Rosenband,
D.E. Pritchard, W.  Ketterle, cond-matter/0104549 (2001).

\noindent
15. L. Ginzburg, L.P. Pitaevskii, Zh. Eksp. Teor. Fiz. {\bf34}, 1240 (1958) [
Sov. Phys. JETP {\bf 7}, 858 (1958)].

\noindent
16. F. Dalfovo and S. Stringari, Phys. Rev. A{\bf 53}, 2477 (1996).

\noindent
17. Yu. Kagan, E.L. Surkov, and G.V. Shlyapnikov, Phys. Rev. A{\bf 54}, R173 (1996).

\noindent 
18.W. Ketterle, D.S. Durfee and D.M. Stamper-Kurn, in {\it Bose-Einstein Condensation in Atomic Gases}, edited by M. Inguscio, S. Stringari and C.E. Wieman (IOS Press, Amsterdam) 1999, p.67.
\noindent
\end{document}